\documentclass{cimento_PN}

\usepackage{graphicx}\usepackage{epsfig}\usepackage{portland}
\usepackage{amsmath}\usepackage{amssymb}\usepackage{amsfonts}
\usepackage{bm}
\usepackage{exscale}\usepackage{rotating}
\usepackage{dcolumn}
\usepackage{color}
\usepackage{enumitem}
\usepackage{soul}
\usepackage{ulem}

\let\oldv\v
\def\diff{\mathrm{d}}

\def\v{\mathrm{v}}

\def\A{\mathrm{A}}
\def\B{\mathrm{B}}
\def\C{\mathrm{C}}
\def\D{\mathrm{D}}

\def\Ek{E_{\textrm{k}}}

\def\vecr{\mathbf{r}}
\def\vecp{\mathbf{p}}

\def\Ntest{N_{\textrm{test}}}

\title{A non-equilibrium microscopic description of spallation}
\author{P.~Napolitani\from{ins:x}\ETC, \atque M.~Colonna\from{ins:y}}
\instlist{
	\inst{ins:x} IPN, CNRS/IN2P3, Univ. Paris-Sud 11, Univ. Paris-Saclay, 91406 Orsay Cedex, France
	\inst{ins:y} INFN-LNS, Laboratori Nazionali del Sud, 95123 Catania, Italy}
\PACSes{\PACSit{24.10.Cn}{Many-body theory}
\PACSit{25.70.Pq}{Multifragment emission and correlations}
\PACSit{25.40.Sc}{Spallation reactions}
}

\begin{document}

\maketitle

\begin{abstract}
	We investigate the prompt emission of few intermediate-mass fragments in spallation reactions induced by protons and deuterons in the 1 GeV range.
	Such emission has a minor contribution to the total reaction cross section, but it may overcome evaporation and fission channels in the formation of light nuclides.
	The role of mean-field dynamics and phase-space fluctuations in these reactions is investigated through the Boltzmann-Langevin transport equation.
    We found that a process of frustrated fragmentation and re-aggregation is a prominent mechanism of production of IMFs which can not be assimilated to the statistical decay of a compound nucleus.
	Very interestingly, this process may yield a small number of IMF in the exit channel, which may even reduce to two, and be wrongly confused with ordinary asymmetric fission.
    This interpretation, inspired by nuclear-spallation experiments, can be generalised to heavy-ion collisions approaching the fragmentation threshold. 
\end{abstract}

\section{Introduction, context, and need for a dynamical approach}
% Historical, contradictory experimental information? Dynamical approach needed.
%
In a 1947 publication, Seaborg~\cite{Seaborg1947} coined the term \textit{spallation} as a nuclear process where the entrance channel is a light high-energy projectile hitting a heavy ion, and the exit channel consists of several particles, where heavy residues or fission fragments combine with light ejectiles, including a large neutron fraction.
	This characteristic production gave rise to several technological applications like neutron sources, secondary beams, hadron therapy and accelerator-driven systems for energy production and transmutation; in astrophysics, spallation contributes to the cosmic-ray isotopic spectrum.
	The same year, Serber~\cite{Serber1947} proposed a schematic description where the process essentially translates in a fast stage of excitation followed by a statistical decay from a fully thermalised source.
	The underlying assumption of equilibrium relies on the expectation that light high-energy projectiles induce negligible mechanical perturbations when traversing intermediate-mass or heavy elements, at variance with nucleus-nucleus collisions at Fermi energies.
	Numerical implementations of such scheme were so successful in describing the largest portion of the production cross section that they are still largely employed nowadays in several nuclear-reaction codes for applications.
	Yet, this picture gives an incomplete description of the intermediate-mass fragments (IMFs) which, in some systems, can extend the spallation yields to the gap between the lightest fission-evaporation residues and the light charged particles, as firstly reported by  Nervik and Serber~\cite{Nervik1954} at intermediate energies and by Robb Grover~\cite{RobbGrover1962} at relativistic energies (It would be too long for this paper to list the decades of experiments that followed).
	Still in the spirit of Serber's scheme, where the exit channel is defined as a function of the excitation energy of the spallation remnant, solutions were proposed where IMF emerge either from a series of successive binary fissions, or from a statistical multifragmentation scenario~\cite{Napolitani2004}.
	Even though these prescriptions could even result rather satisfactory, by construction, they were unsuited for following the process as a function of time, leaving some experimental ambiguities unsolved.
	
%
%	--- FIGURE 1
%
\begin{figure}[b!]	
\begin{center}
	\includegraphics[angle=0, width=.8\textwidth]{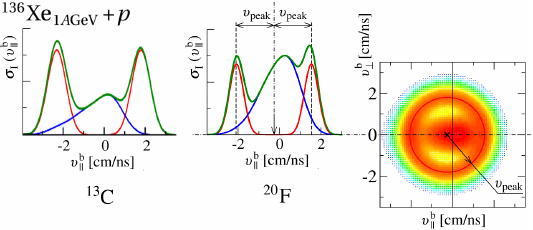}	
\end{center}
\caption{
Experimental zero-angle inclusive velocity spectra for $^{13}$C and $^{20}$F produced in $^{136}$Xe$+p$ at $1A$GeV~\cite{Napolitani2011_2007}, showing two contributions. A perpendicular-versus-parallel-velocity plot illustrates the construction of the zero-angle spectrum for $^{20}$F.
\label{fig1} }
\end{figure}
	To better illustrate the issue, some general considerations may be drawn from the specific example of the system $^{136}$Xe$+p$ at $1A$GeV~\cite{Napolitani2011_2007}.
	Fig.~\ref{fig1} presents zero-angle inclusive velocity spectra of two spallation products, measured at high resolution at the FRagment Separator (GSI, Darmstadt).
	The spectra of all IMF produced in the reaction presented a combination of a concave and a convex contribution, attributed to binary fission and to multifragmentation, respectively, as a result of Coulomb repulsion.
	However, the analysis of event-by-event correlations in a later measurement of the same system~\cite{Gorbinet2012} yielded apparently contradictory experimental information: IMF were found in events with mostly two or three fragments and a large mass asymmetry, i.e. a heavy residue accompanied by one or few IMF, at variance with the traditional multifragmentation scenario where several IMF of equal size are expected.

	In order to address the quest, and understand up to which extent spallation can be included in the multifragmentation picture, it is convenient to drop any a-priori assumption on the degree of equilibration and rather base our investigation on a microscopic dynamical approach: we adopt therefore a transport model.
	A second consideration is that 1$A$GeV proton or deuteron projectiles are sufficient to access density and temperature conditions for collective unstable modes to get amplified: in this case, large-amplitude fluctuations arise, determining the bulk behaviour of the target nucleus, heated up by the interaction with the light projectile, and the corresponding exit channel.
	The grow-rates of unstable modes as a function of the mean-field potential can be efficiently described through a one-body theory.
	Finally, to track a variety of dynamical trajectories in unstable conditions, a stochastic approach is necessary to exploit short-range nucleonic correlations, which are introduced through hard nucleon-nucleon elastic collisions.

\section{BLOB for spallation}
%
%	--- FIGURE 2
%
\begin{figure}[b!]	
\begin{center}
	\includegraphics[angle=0, width=.9\textwidth]{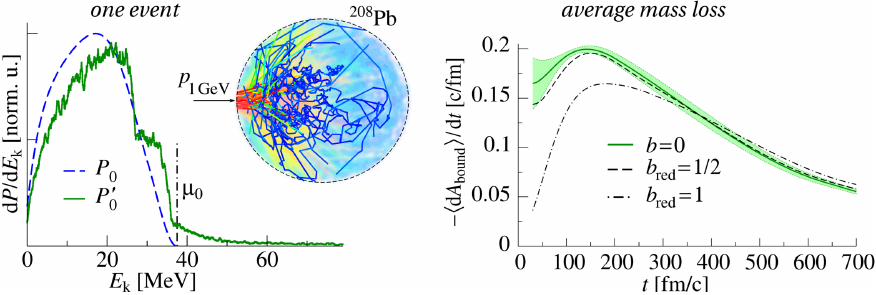}	
\end{center}
\caption{
Left. Kinetic energy distribution of $^{208}$Pb modified by the interaction with a 1 GeV proton. 
Middle. Cascade trajectories and energy-deposition map in configuration space.
Right. average bound mass loss rate as a function of time for three reduced impact parameters $b_{\textrm red}$ (0:central, 1:peripheral); the band indicates the uncertainty for $b=0$.
\label{fig2} }
\end{figure}

	A convenient solution satisfying the requirements listed above is to rely on the Boltzmann-Langevin (BL) equation, represented in terms of one-body distribution function $f(\vecr,\vecp,t)$, i.e. the semiclassical analogue of the Wigner transform of the one-body density matrix, 
\begin{equation}
	\partial_t\,f - \left\{H[f],f\right\} = {\bar{I}[f]}+{\delta I[f]} \;.
\label{eq1}
\end{equation}
The left-hand side gives the evolution of $f$ in its own self-consistent mean field, and the right-hand side represents the residual interaction as composed of an average Boltzmann collision integral $\bar{I}[f]$ and a fluctuating term $\delta I[f]$. 
	$f$ is propagated through the test-particle method and employs a Skyrme-like (SKM*) effective interaction~\cite{Guarnera1996}, defined through a soft isoscalar equation of state (of compressibility $k=200$ MeV) and a linear density dependence of the potential symmetry energy per nucleon.
	The collision term employs a free nucleon-nucleon cross section with a cutoff at 100 mb.
	Compared to a standard Boltzmann-Uehling-Uhlenbeck (BUU) treatment, where the fluctuating term $\delta I[f]$ is neglected, an approximate solution of the BL equation in the spirit of the SMF model~\cite{Colonna1998}, where fluctuations are projected on the coordinate space, is well adapted to describe multifragmentation in heavy-ion collision at Fermi energies.
	Though, in situations where fragment production is largely competing with other mechanisms (fusion in heavy-ion collision and compound nucleus formation in spallation) such approach was found to be not sufficient~\cite{Colonna1997}.

	For spallation reactions and, in general, for situations at the threshold between competing mechanisms, the BL equation should be solved in full phase space.
	The BLOB~\cite{Napolitani2013} approach has been developed for this purpose.
	It replaces the Uehling-Uhlenbeck average collision integral by a treatment which involves extended phase-space agglomerates of test particles of equal isospin A$={a_1,a_2,\dots}$, B$={b_1,b_2,\dots}$ to simulate wave packets (see~\cite{Napolitani2015} for details):
\begin{equation}
	{\bar{I}[f]}+{\delta I[f]}
	= g\int\frac{\diff\vecp_b}{h^3}\,
	\int \diff\Omega\;\;
	W({\scriptstyle\A\B\leftrightarrow\C\D})\;
	F({\scriptstyle\A\B\rightarrow\C\D})\;,
\label{eq2}
\end{equation}
where $W$ is the transition rate, as a function of the nucleon-nucleon cross section divided by $\Ntest$, and $F$ calculates the Pauli blocking probability in terms of occupancy and vacancy of initial and final states for the test-particle agglomerates.	
	A precise shape-modulation technique is applied to ensure that the occupancy distribution does not exceed unity in any phase-space location in the final states.
	At each interval of time, by scanning all phase space in search of collisions, and by redefining all test-particle agglomerates accordingly in phase-space cells of volume $h^3$, nucleon-nucleon correlations are introduced and further exploited within a stochastic procedure.
	Such approach has the advantage of letting fluctuations grow spontaneously while keeping a correct Fermi statistics.

	For applications to spallation reactions, the initial conditions are defined as the heated target nucleus, right after being traversed by the light projectile.
	For this purpose, a simplified cascade treatment can be employed as far as the time spent for the projectile to cross the target is small compared to the dynamical decay process that the target undergoes.
	It consist in letting evolve only the momentum space while keeping the configuration space frozen, till the projectile leaves the target; the energy deposition results from elastic scatterings and partial reflection at potential boundaries.
	As shown in fig.~\ref{fig2}, left, for a $^{208}$Pb target, the initial kinetic energy distribution $P_0(\Ek)$ changes into a new excited configuration $P'_0(\Ek)$, which is used as the initial condition for the dynamics.
	Fig.~\ref{fig2}, right, studies the evolution of the mean number of emitted nucleons per interval of time, i.e. the average mass-loss rate, of the heated remnant as a function of the reduced impact parameter $b_{\textrm red}$.
	Very large $b_{\textrm red}$ would produce a heavy remnant which can be treated through a statistical approach, while central-to-intermediate collisions, with a weak dependence on $b_{\textrm red}$, produce a large mass loss.
	Hereafter we focus on this latter situation.

%\section{Survey of fragment formation}
\section{Frustrated fragmentation and fission by reaggregation}
%
%	--- FIGURE 3
%
\begin{figure}[b!]	
\begin{center}
	\includegraphics[angle=0, width=.9\textwidth]{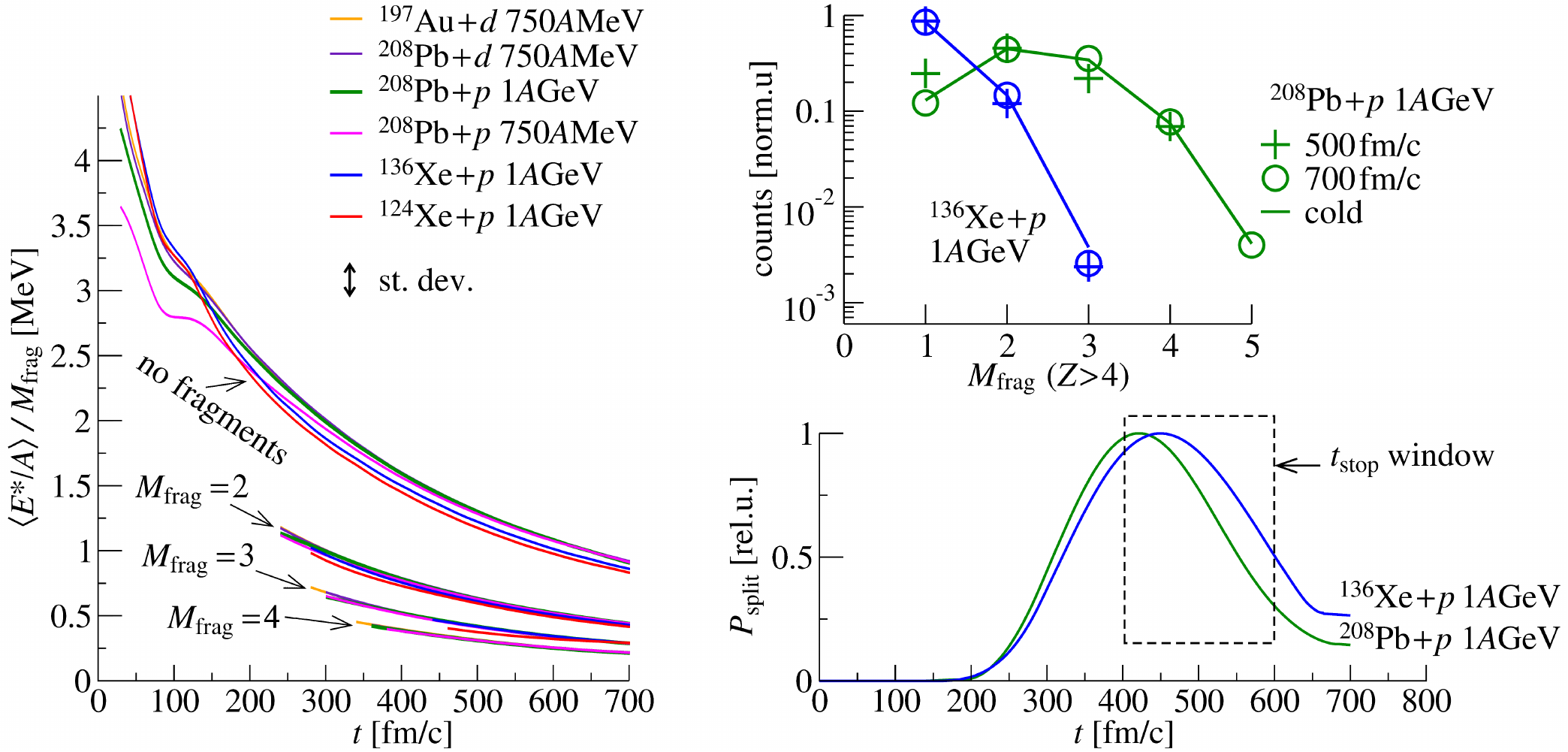}	
\end{center}
\caption{
Survey on fragment formation.
Left. Evolution of the average excitation energy per nucleon for bound matter, divided by the multiplicity of IMF (for better visibility of otherwise overlapping trajectories).
Right-bottom. Probability of separation for a fragment, and window where the last split is favoured.
Right-top. Normalised yields as a function of the IMF multiplicity for $b=0$ (see text).
\label{fig3} }
\end{figure}

	Violent nuclear processes like heavy-ion collisions or spallation are situations where the nuclear system experiences a drop of density and an increase of temperature which may eventually lead to compatible conditions with the spinodal region of the equation of state.
	In these circumstances, unstable modes can develop from fluctuation seeds, and inhomogeneities finally emerge with a size reflecting the wavelength of the leading unstable mode~\cite{Chomaz2004}.
	According to the radial flow induced in the system by the spallation process, these inhomogeneities may become the precursors of fragments which leave the system, or may be reabsorbed while the system reverts to a compact configuration.
	This mechanism is investigated in fig.~\ref{fig3}, left, for a set of spallation systems with different sizes and beam energies.
	In the interval from 50 to 100 fm$/$c, the heavy-nuclear target undergoes an isotropic expansion reaching subsaturation densities and excitation energies exceeding 3 MeV per nucleon.
	These conditions favour the growing of unstable modes so that, later on, inhomogeneities arise in the  bulk and a mottling pattern in configuration space stands out at around 200 fm$/$c.
	In the interval from 200 to 300 fm$/$c, inhomogeneities either combine together into larger blobs of matter, or leave the system as fragments of intermediate mass.
	As shown in the right-bottom panel of fig.~\ref{fig3}, the latest fragments separate in an interval from 400 to 600 fm$/$c, for all the set of systems (only two shown).

	It was found that in a broad interval of time, from the last fragment separation to around 700 fm$/$c, the system reaches a high degree of thermalisation, so that two alternative description coincide closely in describing the evolution of the fragment properties~\cite{Napolitani2015}.
	One description relies on simply progressing further in time within the BLOB approach, while the other consists in switching to a statistical evaporation model.
	For this reason, it is convenient to couple the dynamical transport approach to a statistical evaporation model in this interval of time, in order to continue the decay till the system cools down completely.
	The right-top panel of fig.~\ref{fig3}, shows for two systems the multiplicity of IMF (defined as having $Z>4$) at three times: 500 fm$/$c, corresponding approximately to the last fragment separation, 700 fm$/$c, corresponding to letting the BLOB approach track the process till an excitation energy per nucleon of about 1MeV, and for the cold system, corresponding to a statistical decay afterburner taking over the decay process right after the last fragment separation, event by event.
	It is remarkable to notice that the last two distributions almost correspond, confirming the above discussion.

%
%	--- FIGURE 4
%
\begin{figure}[t!]	
\begin{center}
	\includegraphics[angle=0, width=.99\textwidth]{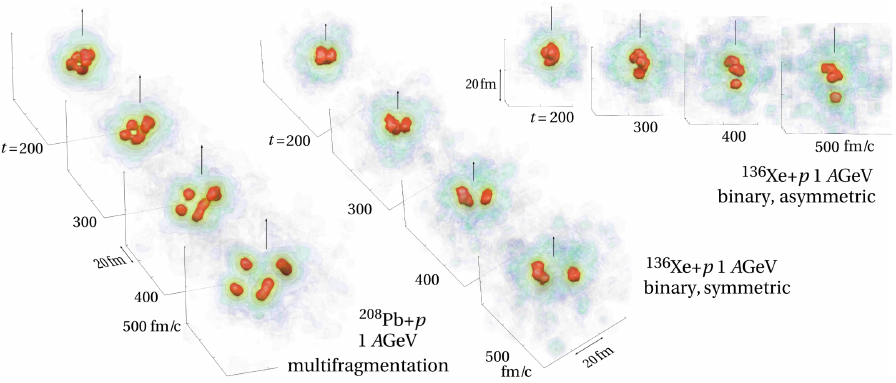}	
\end{center}
\caption{
	Spallation events followed in configuration space and producing fragments through a process of fragmentation and partial reaggregation. 
\label{fig4} }
\end{figure}
	Another important result is the small multiplicity of fragments, which corresponds to a large variety of configurations, including very asymmetric patterns.
	In particular, in heavy system, even though exit channels with more than two IMF are possible, binary split prevail.
	On the other hand, lighter systems mainly revert to one compound nucleus, or undergo binary splits.
	In all these situations, large fragment asymmetries are favoured~\cite{Napolitani2015}.
	The above observations mean that even though density inhomogneities arise rather early and exhibit a spinodal character (almost equal sizes) at around 100fm/c, it takes them long to eventually separate into fragments and this resilience effect smears out the spinodal signal, determining the large probability for asymmetric fragment configurations, binary splits and heavy residues.
	For illustration, three spallation events induced by 1 GeV protons are tracked in configuration space in fig.~\ref{fig4}. 
	The heavy $^{208}$Pb system undergoes a long process where a spinodal pattern gradually rearranges into four fragments which separate at around 300fm$/$c.
	Two events for the $^{136}$Xe systems result into binary splits of different asymmetries.

%
%	--- FIGURE 5
%
\begin{figure}[b!]	
\begin{center}
	\includegraphics[angle=0, width=.7\textwidth]{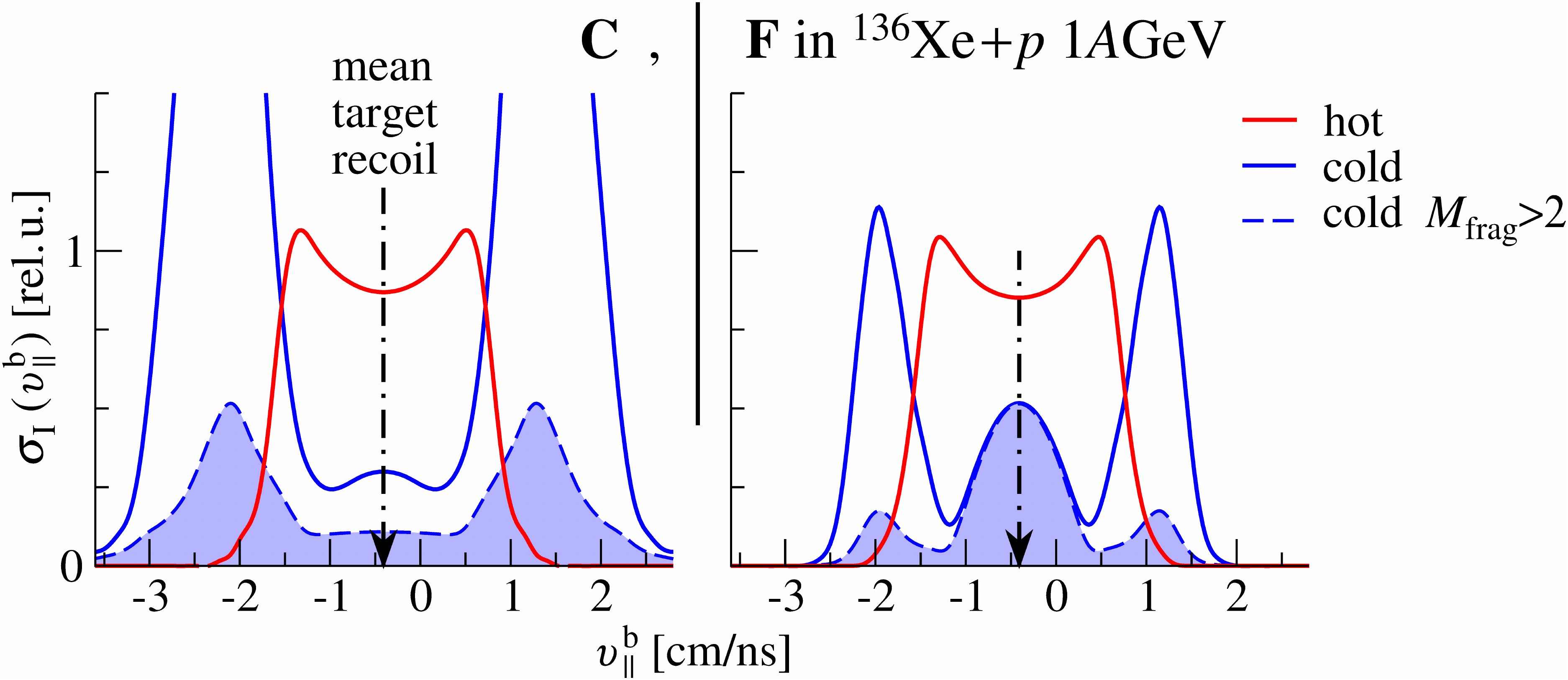}	
\end{center}
\caption{
	Same observable as in fig.~\ref{fig1} obtained from the BLOB approach; hot and cold systems are compared, see text.
\label{fig5} }
\end{figure}

	Finally, fig.~\ref{fig5} retraces kinematic observables similar to fig.~\ref{fig1} for the elements C and F (all isotopes summed up) for $^{136}$Xe$+p$ at 1$A$GeV, with the difference that the calculation can retrace the zero-angle distribution corresponding to the hot system.
	This latter results convex from the folding of the variety of fragment sizes and configurations involved, as expected for a chaotic multifragmentation-like scenario.
	The spectra of the cold system configurations are on the other hand characterised by two wide humps produced from different Coulomb boosts in asymmetric configurations
	Even for contributions corresponding to larger fragment multiplicities than two, the kinematics manifests a binary character due to the size asymmetry.
	The calculation reveals that a binary splits from fragmentation and re-aggregation is a new channel which differs from fission in at least two fundamental aspects.
	Firstly, it develops within the typical short-time chronology of multifragmenation, which can be completly distinguished from the longer times characterising fission processes.
	Secondly, it arises from a mottling low-density phase, without neck formation and without being governed by the potential energy surface.
	In this respect, the discrepancies in the experimental interpretations raised in the introduction of this paper seem solved.

%\section{Residue corridor not attained}
%
%	--- FIGURE 6
%
%\begin{figure}[b!]	
%\begin{center}
%	\includegraphics[angle=0, width=.99\textwidth]{fig6.pdf}	
%\end{center}
%\caption{
%figure .
%\label{fig6} }
%\end{figure}
%

\section{Conclusions}

1$A$GeV light projectiles are sufficient in turning a heavy target into a mechanically unstable system, leading to a variety of asymmetric fragment configurations, ranging from the formation of one heavy residue to multifragmentation, and including binary splits.
	Such competition of mechanisms recalls heavy-ion collisions at Fermi energies, where IMF yields dominate.
Like this latter situation would require, we employed a dynamical description based on a stochastic one-body approach.
	We found that spallation processes at 1$A$GeV may reflect the entrance of the system in the spinodal zone of the equation of state.
	At the same time, the associated behaviour appears frustrated due to the effect of the mean field, which tends to revert the system to a compact shape: the general effect is reducing the IMF multiplicity in the exit channel and imposing large fragment asymmetries.
In particular, as an extreme but very frequent case, we signal the possibility that the system can undergo a process of binary split which can not be assimilated to ordinary asymmetric fission but rather matches a scenario of spinodal fragmentation.
	See refs.~\cite{Napolitani2015, Colonna2016} for further insight and relevant references.

%
% Restore the original definition of the following commands
%
\let\v\oldv

\end{document}